\documentclass[11pt]{article}
\usepackage{amssymb,amsmath,amsfonts}
\usepackage{graphicx}
\usepackage{graphics}
\usepackage{eepic,epsfig}

1.60cm \makeatletter \@addtoreset{equation}{section}

\makeatother

\begin{document}

\title{Scalar Casimir densities induced by a cylindrical \\
shell in de Sitter spacetime}
\author{A. A. Saharian$^{1}$\thanks{%
E-mail: saharian@ysu.am},\thinspace\ V. F. Manukyan$^{2}$ \\
\\
\textit{$^1$Department of Physics, Yerevan State University,}\\
\textit{1 Alex Manoogian Street, 0025 Yerevan, Armenia}\vspace{0.3cm}\\
\textit{$^{2}$Department of Physics and Mathematics, Gyumri State
Pedagogical Institute,}\\
\textit{4 Paruyr Sevak Street, 3126 Gyumri, Armenia}}
\maketitle

\begin{abstract}
We evaluate the positive-frequency Wightman function, the vacuum
expectation values (VEVs) of the field squared and the
energy-momentum tensor for a massive scalar field with general
curvature coupling for a cylindrical shell in background of dS
spacetime. The field is prepared in the Bunch-Davies vacuum state
and on the shell the corresponding operator obeys Robin boundary
condition. In the region inside the shell and for non-Neumann
boundary conditions, the Bunch-Davies vacuum is a physically
realizable state for all values of the mass and curvature coupling
parameter. For both interior and exterior regions, the VEVs are
decomposed into boundary-free dS and shell-induced parts. We show
that the shell-induced part of the vacuum energy-momentum tensor has
a nonzero off-diagonal component corresponding to the energy flux
along the radial direction. Unlike to the case of a shell in
Minkowski bulk, for dS background the axial stresses are not equal
to the energy density. In dependence of the mass and of the
coefficient in the boundary condition, the vacuum energy density and
the energy flux can be either positive or negative. The influence of
the background gravitational field on the boundary-induced effects
is crucial at distances from the shell larger than the dS curvature
scale. In particular, the decay of the VEVs with the distance is
power-law (monotonic or oscillatory with dependence of the mass) for
both massless and massive fields. For Neumann boundary condition the
decay is faster than that for non-Neumann conditions.
\end{abstract}

\bigskip


\bigskip

\section{Introduction}

The study of the Casimir effect (for reviews see \cite{Most97}-\cite{Casi11}%
) for geometries involving cylindrical boundaries have attracted
considerable theoretical and experimental interest. In addition to
traditional problems of quantum electrodynamics under the presence of
material boundaries, the Casimir effect for cylindrical geometries can also
be important to the flux tube models of confinement in quantum
chromodynamics \cite{Fish87,Barb90} and for determining the structure of the
vacuum state in interacting field theories \cite{Ambj83}. A number of widely
used nanostructures, such as single- and multi-walled carbon nanotubes, have
cylindrical shapes. From the point of view of the experimental studies, the
geometries with cylindrical boundaries are among the most optimal candidates
for the precision measurements of the Casimir force. Compared to the case of
spherical boundaries, in these geometries the effective area of interaction
is larger \cite{Brow05}-\cite{Noru12}.

In view of this, the cylindrically symmetric boundary geometries are
becoming increasingly important in the investigations of the Casimir effect.
First the Casimir energy of an infinite perfectly conducting cylindrical
shell has been evaluated in Ref. \cite{Dera81} on the base of a Green's
function technique with an ultraviolet regulator. Later the corresponding
result was rederived by the zeta function technique \cite{Gosd98,Lamb99} and
by using the mode-by-mode summation technique \cite{Milt99} (for the Casimir
energy and self-stresses in a more general problem of a
dielectric-diamagnetic cylinder see \cite{Milt06} and references therein).
The vacuum expectation value (VEV) of the energy-momentum tensor for the
electromagnetic field in the interior and exterior regions of a conducting
cylindrical shell are investigated in \cite{Sah1cyl}. The geometry of two
coaxial cylindrical shells is considered in \cite{Sah2cyl} (see also \cite%
{Sahrev}). The scalar Casimir densities and the vacuum energy for a
single and two coaxial cylindrical shells with Robin boundary
conditions are studied in \cite{Rome01,Saha06}. The zero-point
energy of an arbitrary number of perfectly conducting coaxial
cylindrical shells is calculated in \cite{Tatu08} with the help of
the mode summation technique. Less
symmetric configuration of two eccentric cylinders is considered in \cite%
{Dalv04} by using the mode summation and functional determinant methods. The
Casimir self-energies for an elliptic cylinder are studied in \cite{Kits06}.
The Casimir forces acting on two parallel plates inside a conducting
cylindrical shell are investigated in \cite{Mara07}. The combined geometry
of a wedge and coaxial cylindrical boundary is considered in \cite{Nest01}.
The Casimir interaction energy in the configurations involving cylinders,
plates and spheres has been discussed in \cite{Bord06} (see also \cite%
{Bord09,Casi11}).

In most studies of the Casimir effect with cylindrical boundaries the
geometry of the background spacetime is Minkowskian. Combined effects of a
cylindrical boundary and nontrivial topology induced by a cosmic string are
discussed in \cite{Brev95}. For an idealized infinite straight cosmic string
the spacetime is locally flat except on the top of the string where it has a
delta shaped curvature tensor. In order to see the effects of the curvature
on the Casimir densities induced by a cylindrical boundary, in the present
paper we consider the background geometry described by de Sitter (dS)
spacetime. The corresponding features for planar and spherically-symmetric
boundaries are discussed in \cite{Seta01,Milt12}. The importance of dS
background in gravitational physics is motivated by several reasons. First
of all, dS spacetime is maximally symmetric and better understanding of
physical effects on its backgrounds could serve as a handle to deal with
more general geometries. The investigation of physical effects in dS
spacetime is important for understanding both the early Universe and its
future. In most inflationary scenarios, the dS spacetime is employed to
solve a number of problems in standard cosmology related to initial
conditions in the early Universe. During an inflationary epoch the quantum
fluctuations generate seeds for the formation of large scale structures in
the Universe. More recently, cosmological observations have indicated that
the expansion of the Universe at the present epoch is accelerating and the
corresponding dynamics is well approximated by the model with a positive
cosmological constant as a dominant source. For this source, the standard
cosmology would lead to an asymptotic dS universe in the future.

We have organized the paper as follows. In the next section we evaluate the
positive-frequency Wightman function for a scalar field with general
curvature coupling inside and outside of a cylindrical shell on which the
field obeys Robin boundary condition. We assume that the field is prepared
in the Bunch-Davies vacuum state. The VEV of the field squared is
investigated in section \ref{sec:phi2}. The asymptotics are studied in
detail at distances larger than the dS curvature scale. Section \ref{sec:EMT}
is devoted to the investigation of the VEV of the energy-momentum tensor for
both interior and exterior regions. We show that, in addition to the
diagonal components, the vacuum energy-momentum tensor has an off-diagonal
component which describes an energy flux along the radial direction. The
main results are summarized and discussed in section \ref{sec:Conc}.

\section{Wightman function}

\label{sec:WF}

We consider a quantum scalar field $\varphi \left( x\right) $ in background
of a $\left( D+1\right) $-dimensional dS spacetime, with Robin boundary
condition (BC)
\begin{equation}
(A+Bn^{l}\nabla _{l})\varphi \left( x\right) =0,  \label{BC}
\end{equation}%
imposed on a cylindrical shell having the radius $a$. Here, $n^{l}$ is the
normal to the shell, $\nabla _{l}$ is the covariant derivative operator, $A$
and $B$ are constants. Special cases of (\ref{BC}) correspond to Dirichlet ($%
B=0$) and Neumann ($A=0$) BCs. In accordance with the problem symmetry, we
will write the dS line element in cylindrical coordinates $\left( r,\phi ,%
\mathbf{z}\right) $:%
\begin{equation}
ds^{2}=dt^{2}-e^{2t/\alpha }[dr^{2}+r^{2}d\phi ^{2}+(d\mathbf{z})^{2}],\
\label{ds2}
\end{equation}%
where $\mathbf{z}=(z^{3},...,z^{D})$. The Ricci scalar $R$ and the
corresponding cosmological constant $\Lambda $ are expressed in terms of the
parameter $\alpha $ as
\begin{equation}
R=D(D+1)\alpha ^{-2},\;\Lambda =D(D-1)\alpha ^{-2}/2.  \label{R}
\end{equation}%
In addition to the synchronous time coordinate $t$ it is convenient to
introduce the conformal time in accordance with
\begin{equation}
\tau =-\alpha e^{-t/\alpha },\ -\infty <\tau <0.  \label{eta}
\end{equation}%
The corresponding metric tensor is written in a conformally-flat form, $%
g_{ik}=\Omega ^{2}\eta _{ik}$, with the Minkowskian metric tensor $\eta
_{ik} $ and with the conformal factor $\Omega ^{2}=(\alpha /\tau )^{2}$.

For a free scalar field with a curvature coupling parameter $\xi $ the field
equation is in the form
\begin{equation}
(\nabla _{l}\nabla ^{l}+m^{2}+\xi R)\varphi \left( x\right) =0.  \label{Feq}
\end{equation}%
For the special cases of minimally and conformally coupled fields one has
the values of the curvature coupling $\xi =0$ and $\xi =\xi _{D}=\left(
D-1\right) /(4D)$, respectively. The imposition of boundary condition (\ref%
{BC}) on the field leads to modifications in the vacuum fluctuations
spectrum and, as a result, to the change in the expectation values of the
physical characteristics of the vacuum state $|0\rangle $. For a free field
under consideration all the properties of the vacuum state are encoded in
two-point functions. Here we will investigate the positive-frequency
Wightman function, defined as the VEV $W\left( x,x^{\prime }\right) =\langle
0|\varphi (x)\varphi (x^{\prime })|0\rangle $, assuming that the state $%
|0\rangle $ corresponds to the Bunch-Davies vacuum. Among the set of
maximally symmetric quantum states in dS spacetime, the Bunch-Davies vacuum
is the only one for which the ultraviolet behavior of the two-point
functions is the same as in Minkowski spacetime.

The Wightman function can be presented in the form of the sum over a
complete set of mode functions $\{\varphi _{\sigma }\left( x\right) ,\varphi
_{\sigma }^{\ast }\left( x\right) \}$, obeying the field equation (\ref{Feq}%
) and the boundary condition (\ref{BC}). The collective index $\sigma $ will
be specified below. The Wightman function is given by the expression \
\begin{equation}
W\left( x,x^{\prime }\right) =\sum_{\sigma }\varphi _{\sigma }\left(
x\right) \varphi _{\sigma }^{\ast }\left( x^{\prime }\right) .  \label{WF}
\end{equation}%
Having this function we can evaluate the VEVs of the field squared
and of the energy-momentum tensor. In addition, the Wightman
function determines the transition rate of an Unruh-DeWitt particle
detector in a given state of motion (see, for instance,
\cite{Birr82}).

\subsection{Interior region}

First we consider the region inside the cylindrical shell, $r<a$. In the
cylindrical spatial coordinates, for the corresponding mode functions,
realizing the Bunch-Davies vacuum state, one has
\begin{equation}
\varphi _{\sigma }\left( x\right) =C_{\sigma }\eta ^{D/2}H_{\nu }^{\left(
2\right) }(\gamma \tau )J_{n}\left( \lambda r\right) e^{i\left( n\phi +%
\mathbf{k}\cdot \mathbf{z}\right) },  \label{Modes}
\end{equation}%
where $\eta =|\tau |$, $H_{\nu }^{\left( 2\right) }(x)$ and $J_{n}\left(
x\right) $ are the Hankel and Bessel functions, $\mathbf{k=}\left( k_{3},\
...,\ k_{D}\right) $, $k=\left\vert \mathbf{k}\right\vert $, and we have
defined
\begin{eqnarray}
\nu &=&[D^{2}/4-D\left( D+1\right) \xi -\alpha ^{2}m^{2}]^{1/2},  \notag \\
\gamma &=&\sqrt{\lambda ^{2}+k^{2}}.  \label{nu}
\end{eqnarray}%
With the choice (\ref{Modes}), the collective index $\sigma $ is specified
to $\left( n,\lambda ,\mathbf{k}\right) $, where $n=0,\pm 1,\pm 2,\ldots $,
and $-\infty <k_{l}<+\infty $, $l=3,\ldots ,D$. Note that the parameter $\nu
$ can be either real or purely imaginary. For a conformally coupled massless
field one has $\nu =1/2$ and the Hankel function in (\ref{Modes}) is
expressed in terms of the exponential function. In this case the mode
functions in dS spacetime are related to the corresponding functions for the
shell in Minkowski spacetime by a conformal transformation.

The eigenvalues of the quantum number $\lambda $ are determined from the
boundary condition (\ref{BC}). Substituting the modes (\ref{Modes}), we see
that these eigenvalues are solutions to the equation
\begin{equation}
AJ_{n}\left( \lambda a\right) +B\lambda J_{n}^{\prime }\left( \lambda
a\right) =0,  \label{lambModes}
\end{equation}%
where the prime means the derivative with respect to the argument of
the function. For real $A$ and $B$ the roots of (\ref{lambModes})
are simple and real. We will denote the corresponding positive zeros
by $\lambda a=\lambda _{n,l}$, $l=1,2...$, assuming that they are
arranged in ascending order: $\lambda _{n,l}<\lambda _{n,l+1}$. Now
for the set of quantum numbers specifying the modes one has $\sigma
=(n,l,\mathbf{k})$. Note that for Neumann BC ($A=0$) the zero mode
is present corresponding to $n=0$, $\lambda =0$.

The normalization coefficient $C_{\sigma }$ in (\ref{Modes}) is determined
from the standard condition
\begin{equation}
\int d^{D}x\sqrt{\left\vert g\right\vert }g^{00}\varphi _{\sigma }\left(
x\right) \overleftrightarrow{\partial }_{\tau }\varphi _{\sigma ^{\prime
}}^{\ast }\left( x\right) =i\delta _{\sigma \sigma ^{\prime }},
\label{NormCond}
\end{equation}%
where the integration over the radial coordinate goes over the region inside
the cylinder. In (\ref{NormCond}), $\delta _{\sigma \sigma ^{\prime }}$
stands for the Kronecker delta in the case of discrete components of $\sigma
$ (quantum numbers $n$ and $l$) and for the Dirac delta function for
continuous ones ($\mathbf{k}$). The normalization condition leads to the
result
\begin{equation}
C_{\sigma }^{2}=\dfrac{\lambda T_{n}\left( \lambda a\right) e^{i\left( \nu
^{\ast }-\nu \right) \pi /2}}{4\left( 2\pi \right) ^{D-2}\alpha ^{D-1}a},
\label{C2}
\end{equation}%
with the notation
\begin{equation}
T_{n}\left( z\right) =\dfrac{z}{\left( z^{2}-n^{2}\right) J_{n}^{2}\left(
z\right) +z^{2}J_{n}^{\prime 2}\left( z\right) }.  \label{Tn}
\end{equation}%
In the case of Neumann BC, the normalization coefficient for the zero mode
is obtained from (\ref{C2}) putting $n=0$ and taking the limit $\lambda
\rightarrow 0$.

Substituting the eigenfunctions (\ref{Modes}) into the mode sum formula (\ref%
{WF}) and by taking into account that $H_{\nu }^{(2)}(\gamma \tau )=(2i/\pi
)e^{\nu \pi i/2}K_{\nu }(\gamma \eta e^{-\pi i/2})$, with $K_{\nu }(x)$
being the Macdonald function, for the Wightman function one finds the
expression
\begin{eqnarray}
W\left( x,x^{\prime }\right) &=&\dfrac{4\left( \eta \eta ^{\prime }\right)
^{D/2}}{\left( 2\pi \right) ^{D}a\alpha ^{D-1}}\sum_{n=-\infty }^{\infty
}e^{in\Delta \phi }\int d\mathbf{k}\,e^{i\mathbf{k}\cdot \Delta \mathbf{z}%
}\sum_{l=1}^{\infty }\lambda T_{n}\left( \lambda a\right)  \notag \\
&&\times J_{n}\left( \lambda r\right) J_{n}\left( \lambda r^{\prime }\right)
K_{\nu }(e^{-\pi i/2}\eta \gamma )K_{\nu }(e^{\pi i/2}\eta ^{\prime }\gamma
)|_{\lambda =\lambda _{n,l}/a},  \label{WF2}
\end{eqnarray}%
where $\Delta \phi =\phi -\phi ^{\prime }$, $\Delta \mathbf{z}=\mathbf{z}-%
\mathbf{z}^{\prime }$, and $\gamma $ is defined in (\ref{nu}). For the case
of Neumann BC the contribution of the zero mode should be added to the
right-hand side of (\ref{WF2}). In order to separate explicitly the
contribution induced by the cylindrical shell, we apply to the series over $%
l $ the generalized Abel-Plana summation formula \cite{Saha87,SahaBook}
\begin{eqnarray}
&& 2\sum_{l=1}^{\infty }T_{n}\left( \lambda _{n,l}\right) f\left( \lambda
_{n,l}\right) =\int_{0}^{\infty }dx\,f\left( x\right) +\dfrac{\pi }{2}{%
\mathrm{Res}}{}_{z=0}f\left( z\right) \dfrac{\bar{Y}_{n}\left( z\right) }{%
\bar{J}_{n}\left( z\right) }  \notag \\
&& \qquad -\dfrac{1}{\pi }\int_{0}^{\infty }dx\,\dfrac{\bar{K}_{n}\left(
x\right) }{\bar{I}_{n}\left( x\right) }\left[ e^{-n\pi i}f\left( xe^{\pi
i/2}\right) +e^{n\pi i}f\left( xe^{-\pi i/2}\right) \right] ,
\label{SumForm}
\end{eqnarray}%
where $f\left( z\right) $ is an analytic function on the right half-plane, $%
Y_{n}(z)$ is the Neumann function and, for a given function $F(z)$, we use
the notation
\begin{equation}
\bar{F}\left( z\right) =AF\left( z\right) +(B/a)zF^{\prime }\left( z\right) .
\label{Fbar}
\end{equation}

As the function $f\left( z\right) $ in (\ref{SumForm}) we take
\begin{equation}
f\left( z\right) =zK_{\nu }(e^{-\pi i/2}\eta \sqrt{z^{2}/a^{2}+k^{2}})K_{\nu
}(e^{\pi i/2}\eta ^{\prime }\sqrt{z^{2}/a^{2}+k^{2}})J_{n}\left( zr/a\right)
J_{n}\left( zr^{\prime }/a\right) .  \label{fz}
\end{equation}%
First let us consider the part in the Wightman function corresponding to the
first term in the right-hand side of (\ref{SumForm}). We will denote it by $%
W_{0}\left( x,x^{\prime }\right) $:%
\begin{eqnarray}
W_{0}\left( x,x^{\prime }\right) &=&\dfrac{2\left( \eta \eta ^{\prime
}\right) ^{D/2}}{\left( 2\pi \right) ^{D}\alpha ^{D-1}}\sum_{n=-\infty
}^{\infty }e^{in\Delta \phi }\int d\mathbf{k}\,e^{i\mathbf{k}\cdot \Delta
\mathbf{z}}\int_{0}^{\infty }d\lambda \,\lambda  \notag \\
&&\times J_{n}\left( \lambda r\right) J_{n}\left( \lambda r^{\prime }\right)
K_{\nu }(e^{-\pi i/2}\eta \gamma )K_{\nu }(e^{\pi i/2}\eta ^{\prime }\gamma
).  \label{W0}
\end{eqnarray}%
By using the relations%
\begin{equation}
\sum_{n=-\infty }^{\infty }e^{in\Delta \phi }J_{n}\left( \lambda r\right)
J_{n}\left( \lambda r^{\prime }\right) =J_{0}(\lambda \sqrt{r^{2}+r^{\prime
2}-2rr^{\prime }\cos \Delta \phi }),  \label{W0rel1}
\end{equation}%
and%
\begin{equation}
\int_{0}^{\infty }du\,uF(u)J_{0}(u\sqrt{z_{1}^{2}+z_{2}^{2}})=\frac{1}{2\pi }%
\int_{-\infty }^{+\infty }dk_{1}\int_{-\infty }^{+\infty
}dk_{2}\,e^{ik_{1}z^{1}+ik_{2}z^{2}}F(\sqrt{k_{1}^{2}+k_{2}^{2}}),
\label{W0rel2}
\end{equation}%
one can see that
\begin{equation}
W_{0}\left( x,x^{\prime }\right) =\dfrac{2\left( \eta \eta ^{\prime }\right)
^{D/2}}{\left( 2\pi \right) ^{D+1}\alpha ^{D-1}}\int d\mathbf{K}\,e^{i%
\mathbf{K}\cdot \Delta \mathbf{x}}K_{\nu }(e^{-\pi i/2}\eta |\mathbf{K}%
|)K_{\nu }(e^{\pi i/2}\eta ^{\prime }|\mathbf{K}|),  \label{W01}
\end{equation}%
where $\mathbf{K}=(k_{1},k_{2},k_{3},\ldots ,k_{D})$, $\mathbf{x}%
=(z^{1},z^{2},\mathbf{z})$, $\Delta \mathbf{x}=\mathbf{x}-\mathbf{x}^{\prime
}$, $x=(\tau ,\mathbf{x})$, $x^{\prime }=(\tau ,\mathbf{x}^{\prime })$.
After the evaluation of the integral (see, for instance, \cite{Prud86}), we
get the expression%
\begin{equation}
W_{0}\left( x,x^{\prime }\right) =\frac{\Gamma (D/2+\nu )\Gamma (D/2-\nu )}{%
\left( 4\pi \right) ^{(D+1)/2}\Gamma \left( (D+1)/2\right) \alpha ^{D-1}}%
F\left( \frac{D}{2}+\nu ,\frac{D}{2}-\nu ;\frac{D+1}{2};w\right) ,
\label{W02}
\end{equation}%
where%
\begin{equation}
w=1+\frac{\left( \Delta \eta \right) ^{2}-|\Delta \mathbf{x}|^{2}}{4\eta
\eta ^{\prime }},  \label{w}
\end{equation}%
and $F\left( a,b;c;w\right) $ is the hypergeometric function. The function (%
\ref{W02}) is the Wightman function for the boundary-free dS spacetime.

Note that for $\mathrm{Re}\,\nu \geqslant D/2$ the integral in (\ref{W01})
contains infrared divergences arising from long-wavelength modes. For these
values of $\nu $ the Bunch-Davies vacuum state is not a physically
realizable state in the boundary-free dS spacetime. The boundary conditions
imposed on the field may exclude the modes leading to divergences. This is
the case for the region inside the cylindrical shell with non-Neumann BCs.
For these conditions, in (\ref{WF2}) one has $\gamma \geqslant \lambda
_{n,1}/a$ and the infrared divergences are absent regardless of $\nu $.
Consequently, the Bunch-Davies vacuum is a realizable state for all values
of the parameter $\nu $.

Now, after the application of (\ref{SumForm}) to the series over $l$ in (\ref%
{WF2}), we get the representation
\begin{equation}
W\left( x,x^{\prime }\right) =W_{0}\left( x,x^{\prime }\right) +W_{b}\left(
x,x^{\prime }\right) ,  \label{WFdec}
\end{equation}%
where the contribution induced by the cylindrical shell is given by the
expression
\begin{eqnarray}
W_{b}\left( x,x^{\prime }\right) &=&-\dfrac{4\left( \eta \eta ^{\prime
}\right) ^{D/2}}{\left( 2\pi \right) ^{D}\alpha ^{D-1}}\sideset{}{'}{\sum}%
_{n=0}^{\infty }\cos (n\Delta \phi )\int d\mathbf{k\,}e^{i\mathbf{k}\cdot
\Delta \mathbf{z}}  \notag \\
&&\times \int_{0}^{\infty }du\,u\dfrac{\bar{K}_{n}(ay)}{\bar{I}_{n}(ay)}%
I_{n}(yr)I_{n}(yr^{\prime })\mathcal{I}_{\nu }\left( u\eta ,u\eta ^{\prime
}\right) |_{y=\sqrt{u^{2}+k^{2}}}.  \label{WFb}
\end{eqnarray}%
with the function%
\begin{equation}
\mathcal{I}_{\nu }\left( x,y\right) =I_{-\nu }\left( x\right) K_{\nu }\left(
y\right) +K_{\nu }\left( x\right) I_{\nu }\left( y\right) .  \label{Ical}
\end{equation}%
In (\ref{WFb}), the prime on the sign of the summation means that the term $%
n=0$ is taken with the coefficient 1/2. The formula (\ref{WFb}) is
valid for $\mathrm{Re}\,\nu <1$ and in what follows we assume the
values of $\nu $ in this range. Note that the contribution of the
zero mode for Neumann BC is canceled by the second term in the
right-hand side of (\ref{SumForm}). The
representation (\ref{WFdec}) has two important advantages, compared to (\ref%
{WF2}). First, the explicit knowledge of the roots $\lambda _{n,l}$ is not
required. Second, the effects induced by the shell are explicitly separated
and, for points away from the shell, the boundary-induced contribution is
finite in the coincidence limit of the arguments. In this way, the
renormalization of the VEVs of the field squared and the energy-momentum
tensor is reduced to the one for the boundary-free dS spacetime. In
addition, the integrand in (\ref{WFb}) is an exponentially decreasing
function at the upper limit of the integration instead of strongly
oscillating function in (\ref{WF2}).

\subsection{Exterior region}

In the exterior region, $r>a$, the radial part of the mode functions is a
linear combination of the Bessel and Neumann functions. The relative
coefficient in this combination is determined from the boundary condition (%
\ref{BC}) imposed on the shell. The mode functions realizing the
Bunch-Davies vacuum state are written as
\begin{equation}
\varphi _{\sigma }\left( x\right) =C_{\sigma }\eta ^{D/2}H_{\nu }^{\left(
2\right) }(\tau \gamma )g_{n}\left( \lambda a,\lambda r\right) e^{i\left(
n\phi +\mathbf{k}\cdot \mathbf{z}\right) },  \label{ModesExt}
\end{equation}%
where $0$ $\leqslant \lambda <\infty $ and%
\begin{equation}
g_{n}\left( \lambda a,\lambda r\right) =\bar{Y}_{n}\left( \lambda a\right)
J_{n}\left( \lambda r\right) -\bar{J}_{n}\left( \lambda a\right) Y_{n}\left(
\lambda r\right) ,  \label{gn}
\end{equation}%
with the notation defined by (\ref{Fbar}). Now, in (\ref{NormCond}) the
integration over the radial coordinate goes over the region $a\leqslant
r<\infty $ and in the right-hand side for the part corresponding to the
quantum number $\lambda $ one has $\delta (\lambda -\lambda ^{\prime })$.
For the normalization coefficient we find
\begin{equation}
C_{\sigma }^{2}=\dfrac{\lambda e^{i\left( \nu ^{\ast }-\nu \right) \pi /2}}{%
8\left( 2\pi \right) ^{D-2}\alpha ^{D-1}}\left[ \bar{J}_{n}^{2}\left(
\lambda a\right) +\bar{Y}_{n}^{2}\left( \lambda a\right) \right] ^{-1}.
\label{C2Ext}
\end{equation}

Substituting the functions (\ref{ModesExt}) into the mode sum (\ref{WF}),
and introducing the Macdonald function instead of the Hankel function, we
get the following expression for the exterior Wightman function:
\begin{eqnarray}
W\left( x,x^{\prime }\right) &=&\dfrac{2\left( \eta \eta ^{\prime }\right)
^{D/2}}{\left( 2\pi \right) ^{D}\alpha ^{D-1}}\sum_{n=-\infty }^{\infty
}e^{in\Delta \phi }\int d\mathbf{k}\,e^{i\mathbf{k}\cdot \Delta \mathbf{z}%
}\int_{0}^{\infty }d\lambda \,\lambda  \notag \\
&&\times \dfrac{g_{n}\left( \lambda a,\lambda r\right) g_{n}\left( \lambda
a,\lambda r^{\prime }\right) }{\bar{J}_{n}^{2}\left( \lambda a\right) +\bar{Y%
}_{n}^{2}\left( \lambda a\right) }K_{\nu }(e^{-\pi i/2}\eta \gamma )K_{\nu
}(e^{\pi i/2}\eta ^{\prime }\gamma ).  \label{WFext}
\end{eqnarray}%
By using the identity%
\begin{equation}
\dfrac{g_{n}\left( \lambda a,\lambda r\right) g_{n}\left( \lambda a,\lambda
r^{\prime }\right) }{\bar{J_{n}}^{2}\left( \lambda a\right) +\bar{Y_{n}}%
^{2}\left( \lambda a\right) }=J_{n}\left( \lambda r\right) J_{n}\left(
\lambda r^{\prime }\right) -\dfrac{1}{2}\sum_{j=1,2}\dfrac{\bar{J}_{n}\left(
\lambda a\right) }{\bar{H}_{n}^{\left( j\right) }\left( \lambda a\right) }%
H_{n}^{\left( j\right) }\left( \lambda r\right) H_{n}^{\left( j\right)
}\left( \lambda r^{\prime }\right) ,  \label{Ident}
\end{equation}%
the Wightman function is presented in the decomposed form (\ref{WFdec}) with
the shell-induced part
\begin{eqnarray}
W_{b}\left( x,x^{\prime }\right) &=&-\dfrac{\left( \eta \eta ^{\prime
}\right) ^{D/2}}{\left( 2\pi \right) ^{D}\alpha ^{D-1}}\sum_{n=-\infty
}^{\infty }e^{in\Delta \phi }\int d\mathbf{k}\,e^{i\mathbf{k}\cdot \Delta
\mathbf{z}}\sum_{j=1,2}\int_{0}^{\infty }d\lambda \,\lambda \dfrac{\bar{J}%
_{n}\left( \lambda a\right) }{\bar{H}_{n}^{\left( j\right) }\left( \lambda
a\right) }  \notag \\
&&\times H_{n}^{\left( j\right) }\left( \lambda r\right) H_{n}^{\left(
j\right) }\left( \lambda r^{\prime }\right) K_{\nu }(e^{-\pi i/2}\eta \gamma
)K_{\nu }(e^{\pi i/2}\eta ^{\prime }\gamma ).  \label{WFbext}
\end{eqnarray}

Assuming that the functions $\bar{H}_{n}^{\left( 1\right) }\left( z\right) $
and $\bar{H}_{n}^{\left( 2\right) }\left( z\right) $ have no zeros for $%
0<\arg z\leqslant \pi /2$ and $-\pi /2\leqslant \arg z<0$
respectively, we rotate the contour of integration over $\lambda $
by the angle $\pi /2$ for the term with $j=1$ and by the angle $-\pi
/2$ for $j=2$. The expression (\ref{WFbext}) takes the form
\begin{eqnarray}
W_{b}\left( x,x^{\prime }\right) &=&-\dfrac{4\left( \eta \eta ^{\prime
}\right) ^{D/2}}{\left( 2\pi \right) ^{D}\alpha ^{D-1}}\sideset{}{'}{\sum}%
_{n=0}^{\infty }\cos (n\Delta \phi )\int d\mathbf{k}\,e^{i\mathbf{k}\cdot
\Delta \mathbf{z}}  \notag \\
&&\times \int_{0}^{\infty }du\,\,u\dfrac{\bar{I}_{n}(ay)}{\bar{K}_{n}(ay)}%
K_{n}(ry)K_{n}(r^{\prime }y)\mathcal{I}_{\nu }\left( u\eta ,u\eta ^{\prime
}\right) |_{y=\sqrt{u^{2}+k^{2}}}.  \label{WFbext3}
\end{eqnarray}%
Comparing with (\ref{WFb}), we see that the boundary-induced part of the
Wightman function in the exterior region is obtained from the corresponding
expression in the interior region by the interchange $I_{n}\rightleftarrows
K_{n}$.

The expressions of the Wightman functions inside and outside a cylindrical
shell in Minkowski spacetime are obtained from (\ref{WFb}) and (\ref{WFbext3}%
) in the limit $\alpha \rightarrow \infty $. In order to show that, we note
that for large values of $\alpha $ one has $\nu \approx im\alpha $ and $\eta
\approx \alpha -t$. By using the uniform asymptotic expansions of the
functions $I_{\pm \nu }\left( |\nu |z\right) $ and $K_{\nu }\left( |\nu
|z\right) $ (see \cite{Olve74}) for purely imaginary values of the order
with a large modulus, it can be seen that the main contribution to the
integrals comes from the range $u>m$ in which one has \cite{Milt12}%
\begin{equation}
\mathcal{I}_{\nu }\left( u\eta ,u\eta ^{\prime }\right) \approx \frac{\cosh
(\Delta t\sqrt{u^{2}-m^{2}})}{\alpha \sqrt{u^{2}-m^{2}}},  \label{IcalAs}
\end{equation}%
with $\Delta t=t^{\prime }-t$. Substituting into (\ref{WFb}), for the
Wightman function in the interior region we find
\begin{eqnarray}
W_{M,b}\left( x,x^{\prime }\right) &=&-\dfrac{4}{\left( 2\pi \right) ^{D}}%
\sideset{}{'}{\sum}_{n=0}^{\infty }\cos (n\Delta \phi )\int d\mathbf{k\,}e^{i%
\mathbf{k}\cdot \Delta \mathbf{z}}\int_{\sqrt{k^{2}+m^{2}}}^{\infty }dy\,y
\notag \\
&&\times \frac{\cosh (\Delta t\sqrt{y^{2}-k^{2}-m^{2}})}{\sqrt{%
y^{2}-k^{2}-m^{2}}}\dfrac{\bar{K}_{n}(ay)}{\bar{I}_{n}(ay)}%
I_{n}(yr)I_{n}(yr^{\prime }).  \label{WMin}
\end{eqnarray}%
The expression in the exterior region is obtained by the interchange $%
I_{n}\rightleftarrows K_{n}$. The corresponding VEVs for the both interior
and exterior regions are investigated in \cite{Rome01}.

\section{VEV of the field squared}

\label{sec:phi2}

The VEVs of the field squared and of the energy-momentum tensor are among
the most important characteristics of the vacuum state. The VEV of the field
squared is obtained from the Wightman function by taking the coincidence
limit of the arguments. Similar to the Wightman function, it is presented in
the decomposed form
\begin{equation}
\langle \varphi ^{2}\rangle =\langle \varphi ^{2}\rangle _{0}+\langle
\varphi ^{2}\rangle _{b},  \label{phi2dec}
\end{equation}%
where $\langle \varphi ^{2}\rangle _{0}$ is the VEV in the boundary-free dS
spacetime and the part $\langle \varphi ^{2}\rangle _{b}$ is induced by the
cylindrical shell. The boundary-free part is widely investigated in the
literature. Because of the maximal symmetry of the dS spacetime and of the
Bunch-Davies vacuum state, the renormalized boundary-free VEV does not
depend on the spacetime point. In what follows we will be concerned with the
boundary-induced effects.

\subsection{Interior region}

In the region inside the shell, for the boundary-induced contribution from (%
\ref{WFb}) one has%
\begin{equation}
\langle \varphi ^{2}\rangle _{b}=-\dfrac{A_{D}}{\alpha ^{D-1}}%
\sideset{}{'}{\sum}_{n=0}^{\infty }\int_{0}^{\infty }du\,u^{D-1}\dfrac{\bar{K%
}_{n}(ua/\eta )}{\bar{I}_{n}(ua/\eta )}I_{n}^{2}(ur/\eta )h_{\nu }(u),
\label{phi2b}
\end{equation}%
where
\begin{equation}
A_{D}=\frac{\pi ^{-D/2-1}}{2^{D-3}\Gamma (D/2-1)}.  \label{AD}
\end{equation}%
In (\ref{phi2b}), we have defined the function%
\begin{equation}
h_{\nu }(u)=\int_{0}^{1}ds\,s(1-s^{2})^{D/2-2}f_{\nu }\left( us\right) ,
\label{hnu}
\end{equation}%
with%
\begin{equation}
f_{\nu }\left( y\right) =\left[ I_{-\nu }\left( y\right) +I_{\nu }\left(
y\right) \right] K_{\nu }\left( y\right) .  \label{fnuy}
\end{equation}%
In deriving (\ref{phi2b}), we first integrated over the angular part of $%
\mathbf{k}$ in (\ref{WFb}) and then introduced polar coordinates in the $%
(k,u)$-plane. The integral in (\ref{hnu}) is obtained from the integral over
the polar angle. For points outside the shell, $r<a$, the boundary-induced
part is finite and the renormalization is needed for the boundary-free part
only.

The boundary-induced contribution to the VEV depends on $\eta $, $a$ and $r$
in the form of the ratios $a/\eta $ and $r/\eta $. This property is a
consequence of the maximal symmetry of dS spacetime. By taking into account
that $\alpha a/\eta $ is the proper radius of the cylinder and $\alpha
r/\eta $ is the proper distance from the cylinder axis, we see that $a/\eta $
and $r/\eta $ are the proper radius and the proper distance, measured in
units of the dS curvature scale $\alpha $. The function $f_{\nu }\left(
y\right) $ is positive for $\nu \geqslant 0$. In this case, the part in the
VEV of the field squared induced by the cylindrical shell is negative for
Dirichlet BC and positive for Neumann BC. For purely imaginary values of $%
\nu $ the function $f_{\nu }\left( y\right) $ has no definite sign: it is
positive for large values of the argument and oscillates in the region near $%
y=0$.

For a conformally coupled massless field ($\xi =\xi _{D}$, $m=0$), one has $%
\nu =1/2$ and
\begin{equation}
f_{\nu }\left( y\right) =1/y.  \label{InuConf}
\end{equation}%
In this case, for the function $h_{\nu }(u)$ one has%
\begin{equation}
h_{\nu }(u)=\frac{\sqrt{\pi }\Gamma (D/2-1)}{2\Gamma ((D-1)/2)u},
\label{hnuconf}
\end{equation}%
and from (\ref{phi2b}) we obtain
\begin{equation}
\langle \varphi ^{2}\rangle _{b}=\left( \dfrac{\eta }{\alpha }\right)
^{D-1}\langle \varphi ^{2}\rangle _{M},  \label{phi2bc2}
\end{equation}%
where
\begin{equation}
\langle \varphi ^{2}\rangle _{M}=-\dfrac{2^{2-D}\pi ^{-(D+1)/2}}{\Gamma
((D-1)/2)}\sideset{}{'}{\sum}_{n=0}^{\infty }\int_{0}^{\infty }du\,u^{D-2}%
\dfrac{\bar{K}_{n}(au)}{\bar{I}_{n}(au)}I_{n}^{2}(ru),  \label{phi2M}
\end{equation}%
is the corresponding VEV inside a cylindrical shell in the Minkowski
spacetime.

The boundary-induced part $\langle \varphi ^{2}\rangle _{b}$
diverges on the shell. The surface divergences in local physical
characteristics of the vacuum state are well known in quantum field
theory with boundaries. They are investigated for various types of
bulk and boundary geometries. In the geometry under consideration,
for points near the shell, the dominant contribution to
(\ref{phi2b}) comes from large values of $u$ and $n$. By taking into
account that for large $y$ one has $f_{\nu }\left( y\right) \approx
1/y$, we conclude that the leading term in the asymptotic expansion
over the distance from the boundary coincides with that for a
conformally coupled massless field. By taking into account the
relation (\ref{phi2bc2}) and using the corresponding asymptotic for
the
shell in the Minkowski spacetime, we get%
\begin{equation}
\langle \varphi ^{2}\rangle _{b}\approx -\frac{\Gamma ((D-1)/2)(2\delta
_{0B}-1)}{(4\pi )^{(D+1)/2}[\alpha (a-r)/\eta ]^{D-1}}.  \label{phi2bnear}
\end{equation}%
In deriving (\ref{phi2bnear}) for $B\neq 0$ we have assumed that $a-r\ll |B|$%
. Note that $\alpha (a-r)/\eta $ is the proper distance from the shell. As
it seen, near the shell the boundary-induced part in the VEV of the field
squared is negative for Dirichlet BC ($B=0$) and positive for non-Dirichlet
BC.

On the axis of the shell the contribution of the term with $n=0$ survives
only and we get%
\begin{equation}
\langle \varphi ^{2}\rangle _{b,r=0}=-\dfrac{A_{D}\alpha ^{1-D}}{2(a/\eta
)^{D}}\int_{0}^{\infty }dx\,\,x^{D-1}h_{\nu }(x\eta /a)\dfrac{\bar{K}_{0}(x)%
}{\bar{I}_{0}(x)}.  \label{phi2Axis}
\end{equation}%
This expression is further simplified for large values of $a/\eta $
corresponding to large values of the shell proper radius compared to the dS
curvature radius. By using the asymptotic expressions for the modified
Bessel functions with small values of the argument, to the leading order,
for $u\ll 1$ one gets%
\begin{equation}
h_{\nu }\left( u\right) \approx \frac{\Gamma (D/2-1)}{4}\sigma _{\nu }{%
\mathrm{Re}}\left[ \frac{(2/u)^{2\nu }\Gamma (\nu )}{\Gamma (D/2-\nu )}%
\right] ,  \label{hnuas}
\end{equation}%
where $\sigma _{\nu }=1$ for positive $\nu $ and $\sigma _{\nu }=2$ for
purely imaginary $\nu $. Substituting into (\ref{phi2Axis}), for positive $%
\nu $ we find%
\begin{equation}
\langle \varphi ^{2}\rangle _{b,r=0}\approx -\dfrac{\pi ^{-D/2-1}A^{\mathrm{%
(i)}}(\nu )}{\alpha ^{D-1}(2a/\eta )^{D-2\nu }},  \label{phi2Axis1}
\end{equation}%
where%
\begin{equation}
A^{\mathrm{(i)}}(\nu )=\frac{\Gamma (\nu )}{\Gamma (D/2-\nu )}%
\int_{0}^{\infty }dx\,\,x^{D-2\nu -1}\dfrac{\bar{K}_{0}(x)}{\bar{I}_{0}(x)}.
\label{A0c}
\end{equation}%
In this case $\langle \varphi ^{2}\rangle _{b,r=0}$ is a monotonic function
of $a/\eta $.

For purely imaginary $\nu $ and for $a/\eta \gg 1$, the decay of the leading
term is oscillatory:%
\begin{equation}
\langle \varphi ^{2}\rangle _{b,r=0}\approx -\dfrac{2\pi ^{-D/2-1}M^{\mathrm{%
(i)}}}{\alpha ^{D-1}(2a/\eta )^{D}}\cos [2|\nu |\ln (2a/\eta )+\phi ^{%
\mathrm{(i)}}],  \label{phi2Axis2}
\end{equation}%
where $M^{\mathrm{(i)}}$ and $\phi ^{\mathrm{(i)}}$ are defined by the
relation%
\begin{equation}
A^{\mathrm{(i)}}(\nu )=M^{\mathrm{(i)}}e^{i\phi ^{\mathrm{(i)}}},
\label{M0c}
\end{equation}%
with $M^{\mathrm{(i)}}=|A^{\mathrm{(i)}}(\nu )|$. For a given value of $a$,
the expressions (\ref{phi2Axis1}) and (\ref{phi2Axis2}) describe the
behavior of the VEV at late times of the expansion, $t\gg \alpha $. In the
case of positive $\nu $ the shell-induced VEV on the axis decays as $%
e^{-(D-2\nu )t/\alpha }$, whereas for purely imaginary $\nu $ the decay is
like $e^{-Dt/\alpha }\cos (\omega t+\phi ^{\mathrm{(i)}})$ with $\omega
=2|\nu |\alpha ^{-1}\ln (2a/\eta )$.

\subsection{Exterior region}

In the region outside the cylindrical shell, taking the coincidence limit of
the arguments in (\ref{WFbext3}), for the boundary-induced part in the VEV
of the field squared we get%
\begin{equation}
\langle \varphi ^{2}\rangle _{b}=-\dfrac{A_{D}}{\alpha ^{D-1}}%
\sideset{}{'}{\sum}_{n=0}^{\infty }\int_{0}^{\infty }du\,u^{D-1}\dfrac{\bar{I%
}_{n}(ua/\eta )}{\bar{K}_{n}(ua/\eta )}K_{n}^{2}(ur/\eta )h_{\nu }(u),
\label{phi2ext}
\end{equation}%
with the function $h_{\nu }(u)$ defined in (\ref{hnu}). For $\nu \geqslant 0$
the latter is positive and, similar to the case of the interior region, the
shell-induced VEV is negative for Dirichlet BC and positive for Neumann BC.
The expression in the right-hand side of (\ref{phi2ext})\ diverges on the
shell. The leading term in the asymptotic expansion over the distance from
the shell is given by expression (\ref{phi2bnear}) with $a-r$ replaced by $%
r-a$. In this region the effects of the curvature are small and the leading
term coincides with that for the shell in Minkowski spacetime with the
distance $r-a$ replaced by the proper distance $\alpha (r-a)/\eta $.

At large proper distances from the shell compared with the dS curvature
radius we have $r/\eta \gg 1$ for a fixed value $a/\eta $. In this limit the
dominant contribution to the integral in (\ref{phi2ext}) comes from the
region near the lower limit of the integration, $u\lesssim \eta /r$. For
positive values of $\nu $ and for $A\neq 0$ the dominant contribution to (%
\ref{phi2ext}) comes from the term $n=0$. By taking into account that for
small values of $z$ one has $\bar{I}_{0}(z)/\bar{K}_{0}(z)\approx -1/\ln z$,
to the leading order we find
\begin{equation}
\langle \varphi ^{2}\rangle _{b}\approx -\frac{\pi ^{-(D+1)/2}\alpha
^{1-D}A^{\mathrm{(e)}}(\nu )}{4(2r/\eta )^{D-2\nu }\ln (r/a)},
\label{phi2bFar}
\end{equation}%
where%
\begin{equation}
A^{\mathrm{(e)}}(\nu )=\frac{\Gamma (\nu )\Gamma ^{2}(D/2-\nu )}{\Gamma
(D/2-\nu +1/2)}.  \label{Ae}
\end{equation}%
Here, for $B\neq 0$ we have assumed that $r\gg |B|$. With this condition,
the leading term does not depend on the values of the coefficients in the
boundary condition and is negative. In the case of Neumann BC ($A=0$) and
for positive $\nu $ the leading contribution comes from the terms $n=0$ and $%
n=1$ with the asymptotic%
\begin{equation}
\langle \varphi ^{2}\rangle _{b}\approx \frac{\alpha ^{1-D}A_{N}^{\mathrm{(e)%
}}(\nu )(a/\eta )^{2}}{2\pi ^{(D+1)/2}(2r/\eta )^{D-2\nu +2}},
\label{phi2bFarN}
\end{equation}%
where%
\begin{equation}
A_{N}^{\mathrm{(e)}}(\nu )=\left( 3D/2-3\nu +2\right) \frac{\Gamma (\nu
)\Gamma ^{2}(D/2-\nu +1)}{\Gamma (D/2-\nu +3/2)}.  \label{AeN}
\end{equation}%
For this case the decay of the boundary-induced part al large distances from
the shell is faster and this part is positive. Combining with the asymptotic
analysis for the region near the shell, we conclude that for Robin BC with $%
A,B\neq 0$ the shell-induced contribution in the VEV of the field squared is
positive for points near the shell and negative at large distances. Hence,
for some intermediate value of $r$ it becomes zero. Note that at large
distances the decay of the shell-induced VEV is power-law for both massless
and massive fields. For a cylindrical shell in the Minkowski bulk and for a
massive field the VEV of the field squared decays exponentially with the
distance from the shell.

For purely imaginary values of the parameter $\nu $ and for $A\neq 0$ the
leading asymptotic term is in the form%
\begin{equation}
\langle \varphi ^{2}\rangle _{b}\approx \frac{\pi ^{-(D+1)/2}\alpha ^{1-D}M^{%
\mathrm{(e)}}}{2(2r/\eta )^{D}\ln (a/r)}\cos [2|\nu |\ln (2r/\eta )+\phi ^{%
\mathrm{(e)}}],  \label{phi2bFarIm}
\end{equation}%
where the constants $M^{\mathrm{(e)}}$ and $\phi ^{\mathrm{(e)}}$ are
defined by the relation%
\begin{equation}
M^{\mathrm{(e)}}e^{i\phi ^{\mathrm{(e)}}}=A^{\mathrm{(e)}}(\nu ).  \label{B0}
\end{equation}%
For Neumann BC the asymptotic has the form%
\begin{equation}
\langle \varphi ^{2}\rangle _{b}\approx \frac{\alpha ^{1-D}(a/\eta
)^{2}M_{N}^{\mathrm{(e)}}}{\pi ^{(D+1)/2}(2r/\eta )^{D+2}}\cos [2|\nu |\ln
(2r/\eta )+\phi _{N}^{\mathrm{(e)}}],  \label{phi2bFarNim}
\end{equation}%
with $M_{N}^{\mathrm{(e)}}$ and $\phi _{N}^{\mathrm{(e)}}$ defined as%
\begin{equation}
M_{N}^{\mathrm{(e)}}e^{i\phi _{N}^{\mathrm{(e)}}}=A_{N}^{\mathrm{(e)}}(\nu ).
\label{BN0}
\end{equation}%
As we see, for imaginary $\nu $ the damping of the boundary-induced part
with the distance from the shell is oscillatory.

\section{Energy-momentum tensor}

\label{sec:EMT}

Now we turn to the investigation of the VEV for the energy-momentum
tensor. In addition to describing the physical structure of a
quantum field at a given point, it acts as the source of gravity in
the quasiclassical Einstein equations and plays an important role in
modeling self-consistent dynamics involving the gravitational field.
Similar to the mean field squared, the VEV is decomposed as
\begin{equation}
\langle T_{ik}\rangle =\langle T_{ik}\rangle _{0}+\langle T_{ik}\rangle _{b},
\label{Tik}
\end{equation}%
where $\langle T_{ik}\rangle _{0}$ is the part corresponding to the
boundary-free dS spacetime and $\langle T_{ik}\rangle _{b}$ is the
boundary-induced part. From the maximal symmetry of dS spacetime and of the
Bunch-Davies vacuum state it follows that the renormalized boundary-free
part has the form $\langle T_{i}^{k}\rangle _{0}=\mathrm{const}\cdot \delta
_{i}^{k}$. The boundary-induced contribution is obtained from the
corresponding parts in the Wightman function and in the VEV of the field
squared by using the formula
\begin{equation}
\langle T_{ik}\rangle _{b}=\ \underset{x^{\prime }\rightarrow x}{\lim }%
\partial _{i}\partial _{k}^{\prime }W_{b}\left( x,x^{\prime }\right) +[(\xi
-1/4)g_{ik}\nabla _{l}\nabla ^{l}-\xi \nabla _{i}\nabla _{k}-\xi
R_{ik}]\langle \varphi ^{2}\rangle _{b},  \label{Tikb}
\end{equation}%
with $R_{ik}=Dg_{ik}/\alpha ^{2}$ being the Ricci tensor for dS spacetime.
In the right-hand side of (\ref{Tikb}) we have used the expression for the
energy-momentum tensor of a scalar field which differs from the standard one
(given, for example, in \cite{Birr82}) by a term which vanishes on the
solutions of the field equation (\ref{Feq}) (see \cite{Saha04}).

\subsection{Interior region}

First we consider the region inside the cylindrical shell. After lengthy but
straightforward calculations, the VEVs for the diagonal components are
presented in the form (no summation over $i$)%
\begin{eqnarray}
\langle T_{i}^{i}\rangle _{b} &=&-\dfrac{A_{D}}{\alpha ^{D+1}}%
\sideset{}{'}{\sum}_{n=0}^{\infty }\int_{0}^{\infty }du\,u^{D+1}\dfrac{\bar{K%
}_{n}(ua/\eta )}{\bar{I}_{n}(ua/\eta )}  \notag \\
&&\times \left\{ G_{i}[I_{n}(ur/\eta )]h_{\nu }(u)+I_{n}^{2}(ur/\eta
)\int_{0}^{1}ds\,s^{3}(1-s^{2})^{D/2-2}F_{i}(su)\right\} .  \label{Tiib}
\end{eqnarray}%
In this formula we have defined the functions%
\begin{eqnarray}
G_{0}\left[ f(z)\right] &=&\left( \dfrac{1}{2}-2\xi \right) \left[ f^{\prime
2}(z)+\left( 1+\dfrac{n^{2}}{z^{2}}\right) f^{2}(z)\right] ,  \notag \\
G_{1}\left[ f(z)\right] &=&-\dfrac{f^{\prime 2}(z)}{2}-\dfrac{2\xi }{z}%
f(z)f^{\prime }(z)+\dfrac{1}{2}\left( 1+\dfrac{n^{2}}{z^{2}}\right) f^{2}(z),
\notag \\
G_{2}\left[ f(z)\right] &=&G_{0}\left[ f(z)\right] +\dfrac{2\xi }{z}%
f(z)f^{\prime }(z)-\dfrac{n^{2}}{z^{2}}f^{2}(z)  \notag \\
G_{l}\left[ f(z)\right] &=&G_{0}\left[ f(z)\right] -\dfrac{1}{D-2}f^{2}(z),
\label{Gii}
\end{eqnarray}%
with $l=3,\ldots ,D$, and%
\begin{eqnarray}
F_{0}(z) &=&\left[ \dfrac{1}{4}\partial _{z}^{2}+\left( \dfrac{D+1}{4}-D\xi
\right) \frac{1}{z}\partial _{z}+\frac{m^{2}\alpha ^{2}}{z^{2}}-1\right]
f_{\nu }(z),  \notag \\
F_{1}(z) &=&F_{2}(z)=\left( \xi -\dfrac{1}{4}\right) \partial _{z}^{2}f_{\nu
}(z)+\left[ \xi (D+2)-\dfrac{D+1}{4}\right] \frac{1}{z}\partial _{z}f_{\nu
}(z),  \label{F11} \\
F_{l}(z) &=&F_{1}(z)+\dfrac{1}{D-2}f_{\nu }(z).  \notag
\end{eqnarray}%
Note that, unlike to the case of a shell in Minkowski bulk, here the
stresses along the axial directions are not equal to the energy density.

In addition to the diagonal components, the shell-induced VEV of the
energy-momentum tensor has also a nonzero off-diagonal component
\begin{eqnarray}
\langle T_{0}^{1}\rangle _{b} &=&-\dfrac{A_{D}}{2\alpha ^{D+1}}%
\sideset{}{'}{\sum}_{n=0}^{\infty }\int_{0}^{\infty }du\,u^{D}\,\dfrac{\bar{K%
}_{n}\left( ua/\eta \right) }{\bar{I}_{n}\left( ua/\eta \right) }I_{n}\left(
ur/\eta \right)  \notag \\
&&\times I_{n}^{\prime }\left( ur/\eta \right) \left[ (1-4\xi )u\partial
_{u}+D-4(D+1)\xi \right] h_{\nu }(u).  \label{T01}
\end{eqnarray}%
which corresponds to the energy flux along the radial direction.

The components of the energy-momentum tensor (\ref{Tiib}) and (\ref{T01})
are given in the coordinates $(\tau ,r,\theta ,\varphi )$. For the VEVs in
the coordinates $(t,r,\theta ,\varphi )$ with the synchronous time $t$,
denoted here as $\langle T_{\mathrm{(s)}i}^{k}\rangle _{b}$, one has the
relations (no summation over $i$) $\langle T_{\mathrm{(s)}i}^{i}\rangle
_{b}=\langle T_{i}^{i}\rangle _{b}$ and $\langle T_{\mathrm{(s)}%
0}^{1}\rangle _{b}=(\eta /\alpha )\langle T_{0}^{1}\rangle _{b}$. For a
conformally coupled massless field, by using the relations (\ref{InuConf})
and (\ref{hnuconf}), we can see that the off-diagonal component vanishes and
the diagonal components of the vacuum energy-momentum tensor are related to
the corresponding quantities inside a cylindrical shell in the Minkowski
bulk, given in (\cite{Rome01}), by the conformal relation (no summation over
$i$) $\langle T_{i}^{i}\rangle _{b}\approx (\eta /\alpha )^{D+1}\langle
T_{i}^{i}\rangle _{M}$.

As an additional check of calculations, it can be seen that the
shell-induced VEVs obey the covariant continuity equation, $\nabla
_{k}\langle T_{i}^{k}\rangle _{b}=0$, and the trace relation%
\begin{equation}
\langle T_{i}^{i}\rangle _{b}=[D(\xi -\xi _{D})\nabla _{l}\nabla
^{l}+m^{2}]\langle \varphi ^{2}\rangle _{b}.  \label{TraceRel}
\end{equation}%
In particular, the shell-induced contribution is traceless for a conformally
coupled massless field. The trace anomaly is contained in the boundary-free
part of the VEV. The continuity equation is reduced to two relations between
the components of the shell-induced part:%
\begin{eqnarray}
\left( \partial _{\tau }-\frac{D+1}{\tau }\right) \langle T_{0}^{0}\rangle
_{b}+\frac{1}{\tau }\langle T_{k}^{k}\rangle _{b}+\left( \partial _{r}+\frac{%
1}{r}\right) \langle T_{0}^{1}\rangle _{b} &=&0,  \notag \\
\left( \partial _{\tau }-\frac{D+1}{\tau }\right) \langle T_{1}^{0}\rangle
_{b}+\left( \partial _{r}+\frac{1}{r}\right) \langle T_{1}^{1}\rangle _{b}-%
\frac{1}{r}\langle T_{2}^{2}\rangle _{b} &=&0.  \label{ContRel}
\end{eqnarray}%
Note that $\langle T_{1}^{0}\rangle _{b}=-\langle T_{0}^{1}\rangle _{b}$.

The shell-induced part of the vacuum energy in the region $r\leqslant
r_{0}<a $, per unit coordinate lengths along the directions $z^{3},\ldots
,z^{D}$, is given by
\begin{equation}
E_{r\leqslant r_{0}}^{(b)}=2\pi \left( \alpha /\eta \right)
^{D}\int_{0}^{r_{0}}dr\,r\langle T_{0}^{0}\rangle _{b}.  \label{Ebin}
\end{equation}%
For the corresponding time derivative from the first relation in (\ref%
{ContRel}) one gets%
\begin{equation}
\partial _{t}E_{r\leqslant r_{0}}^{(b)}=\frac{2\pi }{\alpha }\left( \frac{%
\alpha }{\eta }\right) ^{D}\int_{0}^{r_{0}}dr\,r\sum_{l=1}^{D}\langle
T_{l}^{l}\rangle _{b}-2\pi r_{0}\left( \frac{\alpha }{\eta }\right)
^{D-1}\langle T_{0}^{1}\rangle _{b}|_{r=r_{0}}.  \label{Eder}
\end{equation}%
From here it is seen that $\langle T_{0}^{1}\rangle _{b}$ is the energy flux
per unit proper surface area. Note that $\langle T_{l}^{l}\rangle _{b}$ is
the boundary-induced part of the vacuum pressure along the $l$-th direction.
Equation (\ref{Eder}) shows that the change of the energy is caused by two
factors: by the work done by the surrounding (first term in the right-hand
side of (\ref{Eder})) and by the energy flux through the boundary of the
selected volume (second term).

Now let us discuss the asymptotics of the vacuum energy-momentum tensor.
Near the cylindrical surface the dominant contribution to the
boundary-induced VEVs comes from large values of $n$ and $u$. By using the
uniform asymptotic expansions for the modified Bessel functions (see, for
instance, \cite{Abra72}), it can be seen that the leading terms in the
diagonal components for a scalar field with non-conformal coupling ( $\xi
\neq \xi _{D}$) are related to the corresponding terms for a cylindrical
boundary in Minkowski spacetime by (no summation over $i$ ) $\langle
T_{i}^{i}\rangle _{b}\approx (\eta /\alpha )^{D+1}\langle T_{i}^{i}\rangle
_{M}$. These leading terms are given by the expression (no summation over $i$%
)%
\begin{equation}
\langle T_{i}^{i}\rangle _{b}\approx \dfrac{D\Gamma ((D+1)/2)(\xi -\xi _{D})%
}{2^{D}\pi ^{(D+1)/2}[\alpha (a-r)/\eta ]^{D+1}}(2\delta _{B0}-1),
\label{Tiinear}
\end{equation}%
for the components with $i$ $=0,2,...,D$. For the radial stress and the
energy flux, to the leading order, one has%
\begin{equation}
\langle T_{1}^{1}\rangle _{b}\approx \frac{1-r/a}{D}\langle T_{0}^{0}\rangle
_{b},\;\langle T_{0}^{1}\rangle _{b}\approx \frac{a-r}{\eta }\langle
T_{0}^{0}\rangle _{b}.  \label{T11near}
\end{equation}%
The leading terms have opposite signs for Dirichlet and non-Dirichlet BCs.
In particular, for a minimally coupled field the energy density and the
energy flux are negative for Dirichlet BC and positive for non-Dirichlet BC.
Near the shell, the VEVs are dominated by the boundary-induced parts and the
same is the case for the total energy density.

On the axis of the shell, $r=0$, the only nonzero contribution to the
diagonal components of the boundary-induced VEV comes from the terms in (\ref%
{Tiib}) with $n=0,1$. The energy flux vanishes on the axis as $r/\eta $.
Simple expressions on the axis are obtained for large values of the shell
proper radius compared with the dS curvature scale, $a/\eta \gg 1$. For
positive values of $\nu $, to the leading order we have (no summation over $%
i $)%
\begin{equation}
\langle T_{i}^{i}\rangle _{b,r=0}\approx -\dfrac{\pi ^{-D/2-1}A^{\mathrm{(i)}%
}(\nu )}{\alpha ^{D+1}(2a/\eta )^{D-2\nu }}A_{i}^{i},  \label{Tiiaxis}
\end{equation}%
where $A^{\mathrm{(i)}}(\nu )$ is defined by (\ref{A0c}) and
\begin{eqnarray}
A_{0}^{0} &=&D\left[ \xi \left( 2\nu -D-1\right) +\frac{D-2\nu }{4}\right] ,
\notag \\
A_{l}^{l} &=&\frac{2\nu }{D}A_{0}^{0},\;l=1,\ldots ,D.  \label{A00}
\end{eqnarray}%
For a conformally coupled field one has $A_{0}^{0}=(1-2\nu )/4$ and the
leading term (\ref{Tiiaxis}) vanishes in the massless case. For minimally
and conformally coupled massive fields $A_{0}^{0}>0$. Now, by taking into
account that $A^{\mathrm{(i)}}(\nu )>0$ for Dirichlet BC and $A^{\mathrm{(i)}%
}(\nu )<0$ for Neumann BC, we conclude that in these cases $\langle
T_{i}^{i}\rangle _{b,r=0}<0$ for Dirichlet BC and $\langle T_{i}^{i}\rangle
_{b,r=0}<0$ for Neumann BC.

For the energy flux in the limit $r\rightarrow 0$ and for $a/\eta \gg 1$, to
the leading order one has%
\begin{equation}
\langle T_{0}^{1}\rangle _{b}\approx -\dfrac{4\pi ^{-D/2-1}A_{10}^{\mathrm{%
(i)}}(\nu )r/\eta }{\alpha ^{D+1}(2a/\eta )^{D-2\nu +2}},  \label{T01axis}
\end{equation}%
with the function%
\begin{equation}
A_{10}^{\mathrm{(i)}}(\nu )=\frac{\Gamma (\nu )A_{0}^{0}}{D\Gamma (D/2-\nu )}%
\int_{0}^{\infty }dx\,x^{D-2\nu +1}\,\left[ \dfrac{\bar{K}_{0}\left(
x\right) }{\bar{I}_{0}\left( x\right) }+\dfrac{\bar{K}_{1}\left( x\right) }{%
\bar{I}_{1}\left( x\right) }\right] .  \label{A01}
\end{equation}%
As before, for a conformally coupled massless field the leading term
vanishes. For minimally and conformally coupled massive fields the
energy flux corresponding to (\ref{T01axis}) is negative for
Dirichlet BC and positive for Neumann BC.

For purely imaginary $\nu $ and for $a/\eta \gg 1$, the behavior of the
diagonal components on the axis is described by (no summation over $i$)
\begin{equation}
\langle T_{i}^{i}\rangle _{b,r=0}\approx -\dfrac{2\pi ^{-D/2-1}M_{i}^{%
\mathrm{(i)}}}{\alpha ^{D+1}(2a/\eta )^{D}}\cos [2|\nu |\ln (2a/\eta )+\phi
_{i}^{\mathrm{(i)}}],  \label{Tiiaxis2}
\end{equation}%
with $M_{i}^{\mathrm{(i)}}=|A_{i}^{\mathrm{(i)}}(\nu )A_{i}^{i}|$ and the
phase $\phi _{i}^{\mathrm{(i)}}$ defined by the relation
\begin{equation}
A_{i}^{\mathrm{(i)}}(\nu )A_{i}^{i}=M_{i}^{\mathrm{(i)}}e^{i\phi _{i}^{%
\mathrm{(i)}}}.  \label{Mic}
\end{equation}%
For the energy flux near the axis, in the case of imaginary $\nu $ one has
the leading term%
\begin{equation}
\langle T_{0}^{1}\rangle _{b}\approx -\dfrac{8\pi ^{-D/2-1}M_{10}^{\mathrm{%
(i)}}r/\eta }{\alpha ^{D+1}(2a/\eta )^{D+2}}\cos [2|\nu |\ln (2a/\eta )+\phi
_{10}^{\mathrm{(i)}}],  \label{T10axis1}
\end{equation}%
where%
\begin{equation}
A_{10}^{\mathrm{(i)}}(\nu )=M_{10}^{\mathrm{(i)}}e^{i\phi _{10}^{\mathrm{(i)}%
}}.  \label{MC10}
\end{equation}%
For a given $a$, the expressions (\ref{Tiiaxis}), (\ref{T01axis}), (\ref%
{Tiiaxis2}), (\ref{T10axis1}) describe the asymptotic behavior of the
shell-induced VEVs at late stages of the expansion, $t\gg \alpha $.

\subsection{Exterior region}

Now we turn to the investigation of the VEV for the energy-momentum tensor
outside the cylindrical shell. The VEV is presented in the decomposed form (%
\ref{Tik}) with the diagonal components of the boundary-induced part (no
summation over $i$ )%
\begin{eqnarray}
\langle T_{i}^{i}\rangle _{b} &=&-\dfrac{A_{D}}{\alpha ^{D+1}}%
\sideset{}{'}{\sum}_{n=0}^{\infty }\int_{0}^{\infty }du\,u^{D+1}\dfrac{\bar{I%
}_{n}(ua/\eta )}{\bar{K}_{n}(ua/\eta )}  \notag \\
&&\times \left\{ G_{i}[K_{n}(ur/\eta )]h_{\nu }(u)+K_{n}^{2}(ur/\eta
)\int_{0}^{1}ds\,(1-s^{2})^{D/2-2}s^{3}F_{i}(su)\right\} ,  \label{Tiibext}
\end{eqnarray}%
where the functions $G_{i}[f(z)]$ and $F_{i}(z)$ are defined by (\ref{Gii})
and (\ref{F11}). The off-diagonal component, corresponding to the energy
flux along the radial direction, has the form%
\begin{eqnarray}
\langle T_{0}^{1}\rangle _{b} &=&-\dfrac{A_{D}}{2\alpha ^{D+1}}%
\sideset{}{'}{\sum}_{n=0}^{\infty }\int_{0}^{\infty }du\,u^{D}\,\dfrac{\bar{I%
}_{n}(ua/\eta )}{\bar{K}_{n}(ua/\eta )}K_{n}\left( ur/\eta \right)  \notag \\
&&\times K_{n}^{\prime }\left( ur/\eta \right) \left[ (1-4\xi )u\partial
_{u}+D-4(D+1)\xi \right] h_{\nu }(u).  \label{T01bext}
\end{eqnarray}%
Similar to the case of the interior region, for the components of the vacuum
energy-momentum tensor we have the relations (\ref{TraceRel}) and (\ref%
{ContRel}).

On the boundary the VEVs diverge. The leading terms in the
asymptotic expansion over the distance from the shell for the components $%
\langle T_{i}^{i}\rangle _{b}$ with $i=0,2,\ldots ,D$, are obtained from (%
\ref{Tiinear}) by the replacement $(a-r)\rightarrow (r-a)$. Hence, these
components have the same sign on both sides of the shell for points near the
boundary. The leading terms for the radial stress, $\langle T_{1}^{1}\rangle
_{b}$, and the energy flux, $\langle T_{0}^{1}\rangle _{b}$, are related to
the energy density by (\ref{T11near}). In particular, near the shell the
radial stress and the energy flux for non-conformally coupled fields have
opposite signs in the exterior and interior regions. For a minimally coupled
field, the energy density near the shell is negative for Dirichlet BC and
positive for non-Dirichlet BC ($B\neq 0$). Near the boundary, for both
interior and exterior regions, the corresponding energy flux is directed
from the boundary for Dirichlet BC and to the boundary for non-Dirichlet BC.

Now we consider the asymptotic behavior of VEV for the energy-momentum
tensor at large distances from the cylindrical shell , $r\gg a$. For the
diagonal components the dominant contribution comes from the second term in
the figure braces of (\ref{Tiibext}). For the functions $F_{i}(z)$ in this
term, for small values of the argument one has (no summation over $i$)%
\begin{equation}
F_{i}(z)\approx \sigma _{\nu }{\mathrm{Re}}\left[ \dfrac{2^{2\nu -1}\Gamma
(\nu )}{\Gamma (1-\nu )}A_{i}^{i}z^{-2\nu -2}\right] ,  \label{Fias}
\end{equation}%
with $A_{i}^{i}$ defined by (\ref{A00}). By taking into account that at
large distances from the shell the main contribution to the integrals in (%
\ref{Tiibext}) and (\ref{T01bext}) comes from the region near the lower
limit of the integral, for positive values of $\nu $ and for $A\neq 0$, to
the leading order we get%
\begin{equation}
\langle T_{i}^{k}\rangle _{b}\approx -\dfrac{\pi ^{-(D+1)/2}\alpha
^{-D-1}A_{i}^{k}A^{\mathrm{(e)}}(\nu )}{2^{1+\delta _{i}^{k}}(2r/\eta
)^{D-2\nu +1-\delta _{i}^{k}}\ln (r/a)},  \label{Tikfar}
\end{equation}%
where $A^{\mathrm{(e)}}(\nu )$ is given by (\ref{Ae}) and%
\begin{equation}
A_{0}^{1}=\left( 2\nu /D-1\right) A_{0}^{0}.  \label{A10}
\end{equation}%
The leading term (\ref{Tikfar}) does not depend on the specific value of the
ratio $B/A$. As is seen, at large distances, to the leading order, the
shell-induced stresses are isotropic. For a conformally coupled massless
field $A_{0}^{0}=0$ and the leading term given by (\ref{Tikfar}) vanishes.
In this case the diagonal components decay as $1/r^{D+2}$. For minimally and
conformally coupled massive fields $A_{0}^{0}>0$ and the boundary-induced
part in the energy density is negative at large distances. In accordance
with (\ref{A10}), $\langle T_{0}^{1}\rangle _{b}>0$ and the energy flux is
directed from the shell. For a cylindrical shell in Minkowski spacetime and
for a massless field, at large distances the diagonal components decay as $%
r^{-D}/\ln (r/a)$ and the leading terms vanish for a conformally coupled
field. For massive fields the decay of the Minkowskian VEVs with the
distance from the shell is exponential.

For Neumann BC ($A=0$) and for positive values of the parameter $\nu
$, at large distances, to the leading order, one has
\begin{equation}
\langle T_{i}^{k}\rangle _{b}\approx \dfrac{\pi ^{-(D+1)/2}A_{Ni}^{k}A_{N}^{%
\mathrm{(e)}}(\nu )(a/\eta )^{2}}{2\alpha ^{D+1}(2r/\eta )^{D-2\nu +3-\delta
_{i}^{k}}},  \label{TikfarN}
\end{equation}%
where $A_{Ni}^{i}=A_{i}^{i}$ (no summation over $i$) and
\begin{equation}
A_{N0}^{1}=-2\left( D-2\nu +2\right) A_{0}^{0}/D.  \label{A1N0}
\end{equation}%
As before, for a conformally coupled massless field the leading term
vanishes. For minimally and conformally coupled massive fields the
boundary-induced part in the energy density is positive. In these cases $%
\langle T_{0}^{1}\rangle _{b}<0$ and the energy flux is directed toward the
shell.

Combining with the results of the asymptotic analysis for the region near
the shell, we conclude that for positive values of $\nu $ and for minimally
and conformally coupled massive fields the shell-induced contribution in the
VEV of the energy density is negative/positive near the shell and at large
distances for Dirichlet/Neumann BC. For Robin BC with $A,B\neq 0$ this
contribution is positive near the shell and negative at large distances. The
energy flux is positive/negative for Dirichlet/Neumann BC near the shell and
at large distances. For Robin BC with $A,B\neq 0$ the energy flux is
negative near the shell and positive at large distances. At some
intermediate value the energy flux vanishes.

Asymptotic behavior of the vacuum energy-momentum tensor at large distances
from the shell is qualitatively different for imaginary values of $\nu $. In
this case and for non-Neumann BC ($A\neq 0$) the leading term has the form%
\begin{equation}
\langle T_{i}^{k}\rangle _{b}\approx -\dfrac{\pi ^{-(D+1)/2}\alpha
^{-D-1}M_{i}^{\mathrm{(e)}k}}{2^{\delta _{i}^{k}}(2r/\eta )^{D+1-\delta
_{i}^{k}}\ln (r/a)}\cos [2|\nu |\ln (2r/\eta )+\phi _{i}^{\mathrm{(e)}k}],
\label{Tikfarosc}
\end{equation}%
where the coefficient $M_{i}^{\mathrm{(e)}k}$ and the phase $\phi _{i}^{%
\mathrm{(e)}k}$ are defined by the relation%
\begin{equation}
M_{i}^{\mathrm{(e)}k}e^{i\phi _{i}^{\mathrm{(e)}k}}=A_{i}^{k}A^{\mathrm{(e)}%
}(\nu ).  \label{Cik}
\end{equation}%
The decay of the VEVs is oscillatory. For Neumann BC and in the case of
imaginary $\nu $ for the leading term in the asymptotic expansion one has%
\begin{equation}
\langle T_{i}^{k}\rangle _{b}\approx \dfrac{2\alpha ^{-D-1}M_{Ni}^{\mathrm{%
(e)}k}(a/\eta )^{2}}{\pi ^{(D+1)/2}(2r/\eta )^{D+3-\delta _{i}^{k}}}\cos
[2|\nu |\ln (2r/\eta )+\phi _{Ni}^{\mathrm{(e)}k}],  \label{TikfarNosc}
\end{equation}%
where%
\begin{equation}
M_{Ni}^{\mathrm{(e)}k}e^{i\phi _{Ni}^{\mathrm{(e)}k}}=A_{Ni}^{k}A_{N}^{%
\mathrm{(e)}}(\nu ).  \label{CikN}
\end{equation}%
As is seen, for Neumann BC the suppression of the VEVs at large
distances is faster. For a fixed value of $r$, the asymptotic
expressions given above describe the behavior of the shell-induced
contributions to the VEV of the
energy-momentum tensor at late stages of the expansion corresponding to $%
t\gg \alpha $.

\begin{figure}[tbph]
\begin{center}
\begin{tabular}{cc}
\epsfig{figure=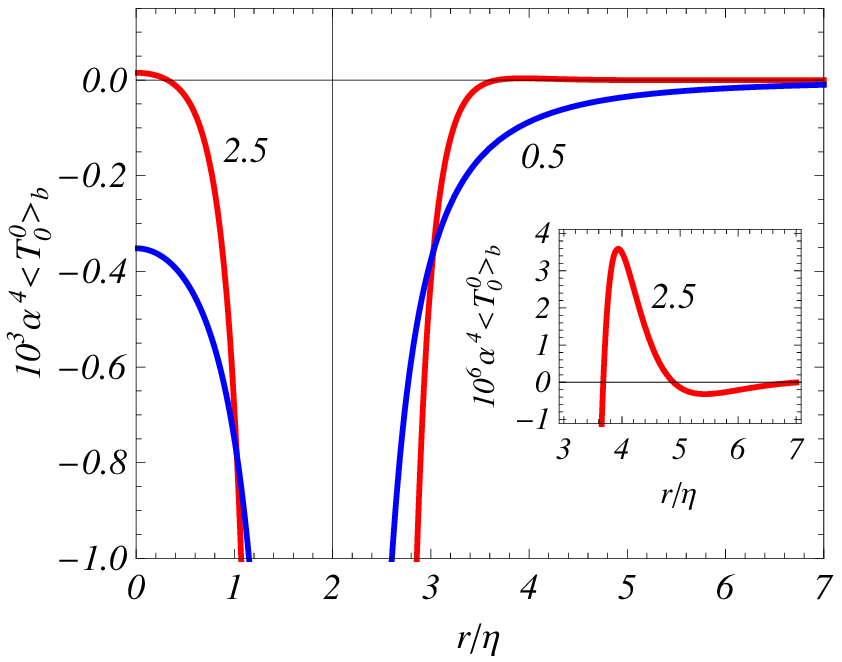,width=7.cm,height=6.cm} & \quad %
\epsfig{figure=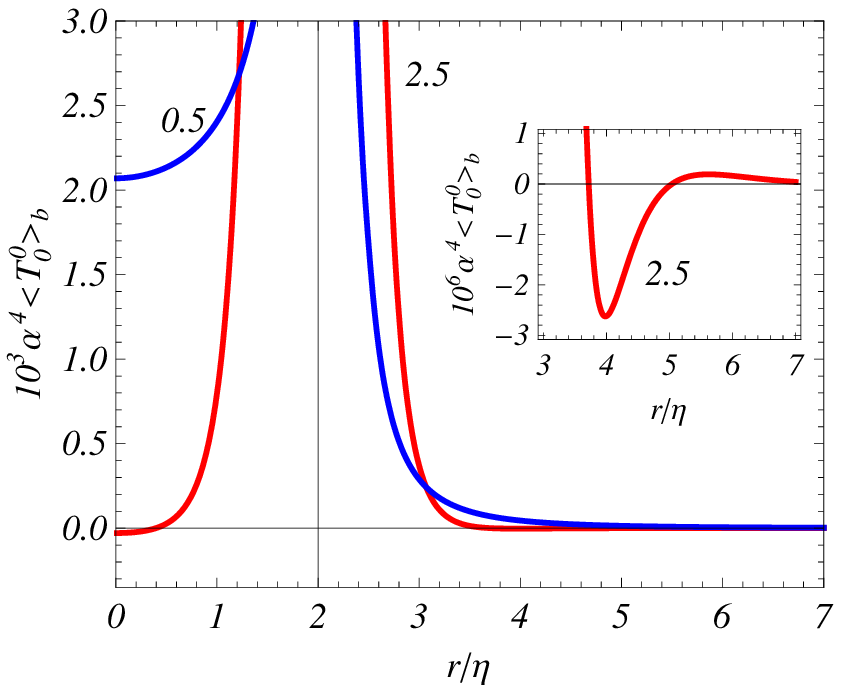,width=7.cm,height=6.cm}%
\end{tabular}%
\end{center}
\caption{Shell-induced contribution in the VEV of the energy density
as a function of the proper distance from the axis in the case of a
conformally coupled field in $D=3$ spatial dimensions. The numbers
near the curves are the values of $m\protect\alpha $ and we have
taken $a/\protect\eta =2$. The left/right panels correspond to
Dirichlet/Neumann BCs.} \label{fig1}
\end{figure}

In figure \ref{fig1}, for a $D=3$ conformally coupled field, we
display the boundary-induced part in the energy density as a
function of the proper distance from the cylindrical shell axis
(measured in units of the dS curvature scale $\alpha $). The graphs
are plotted for the radius of the shell corresponding to $a/\eta =2$
and the numbers near the curves are the values of the parameter
$m\alpha $. The left/right panel corresponds to Dirichlet/Neumann
BC. For $m\alpha =2.5$ the parameter $\nu $ is purely imaginary and
the oscillatory behavior at large distances from the shell is seen.
Similar to the minimal coupling, for a conformally coupled field the
energy density near the shell is negative for Dirichlet BC and
positive for Neumann BC.

\begin{figure}[tbph]
\begin{center}
\begin{tabular}{cc}
\epsfig{figure=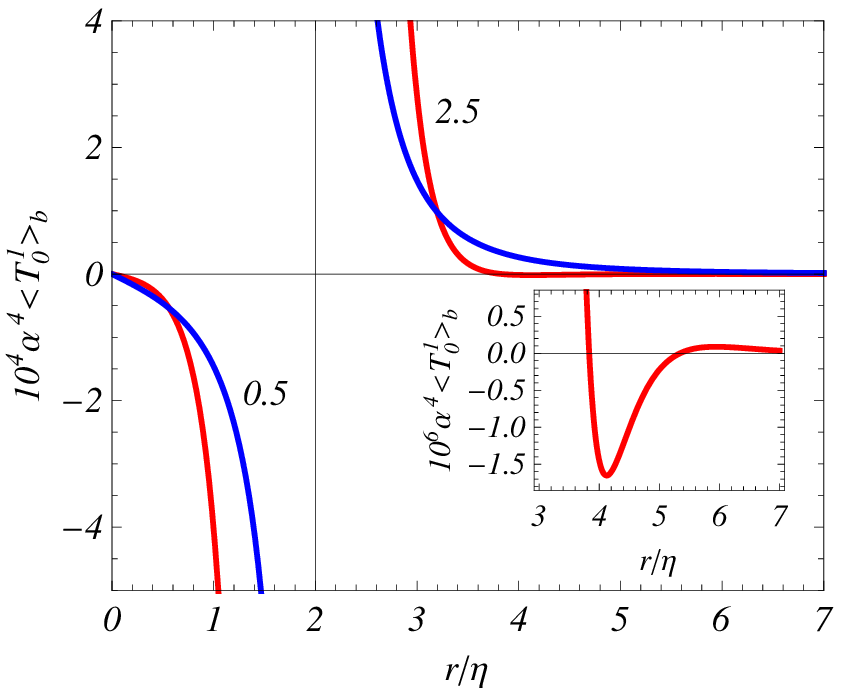,width=7.cm,height=6.cm} & \quad %
\epsfig{figure=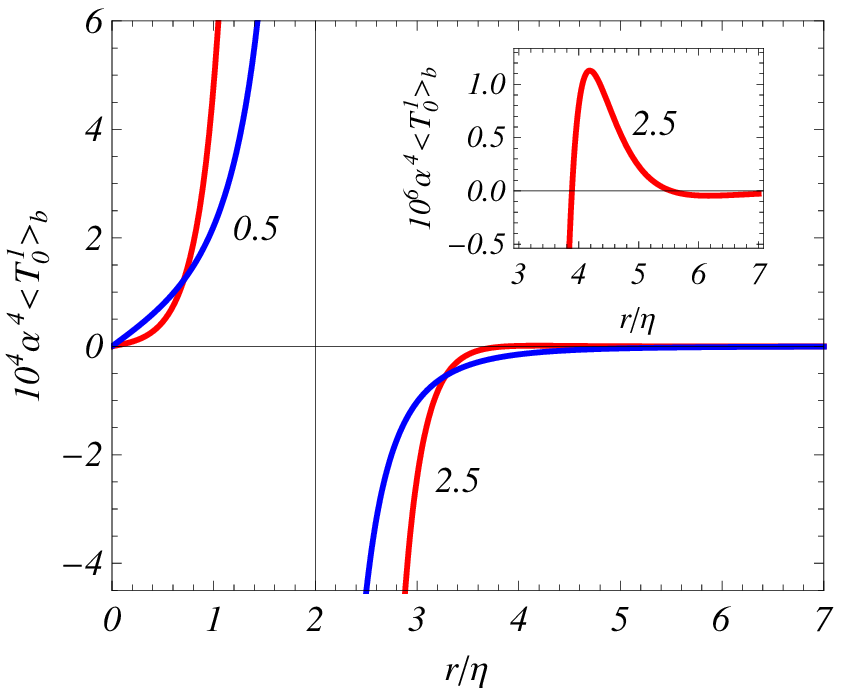,width=7.cm,height=6.cm}%
\end{tabular}%
\end{center}
\caption{The same as in figure \protect\ref{fig1} for the energy flux.}
\label{fig2}
\end{figure}

The same graphs for the energy flux are plotted in figure \ref{fig2}. Inside
the shell, the energy flux is negative for Dirichlet BC and positive for
Neumann BC. This means that the flux is directed from the shell for the
first case and toward the shell in the second case. On the shell axis the
energy flux vanishes as $r/\eta $.

\begin{figure}[tbph]
\begin{center}
\begin{tabular}{cc}
\epsfig{figure=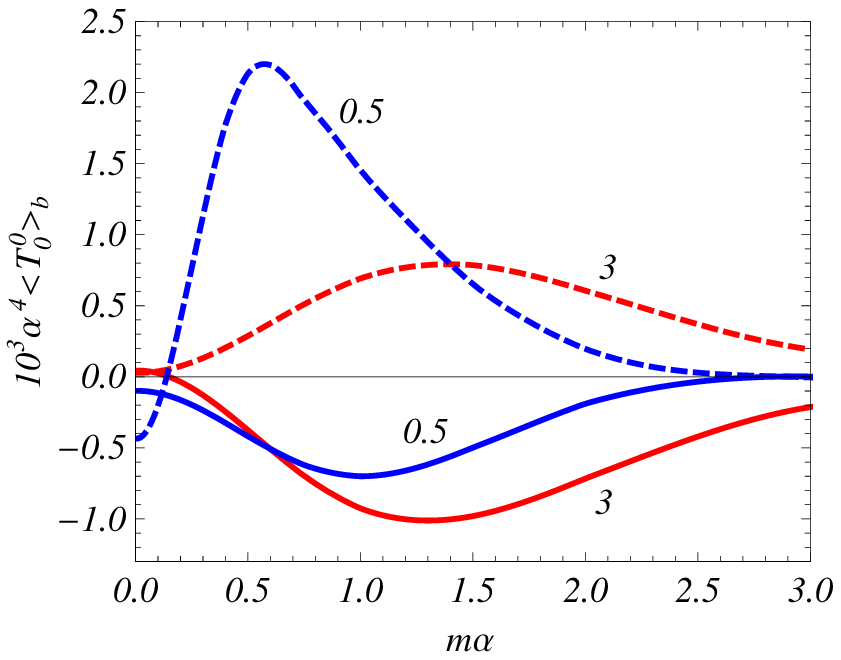,width=7.cm,height=6.cm} & \quad %
\epsfig{figure=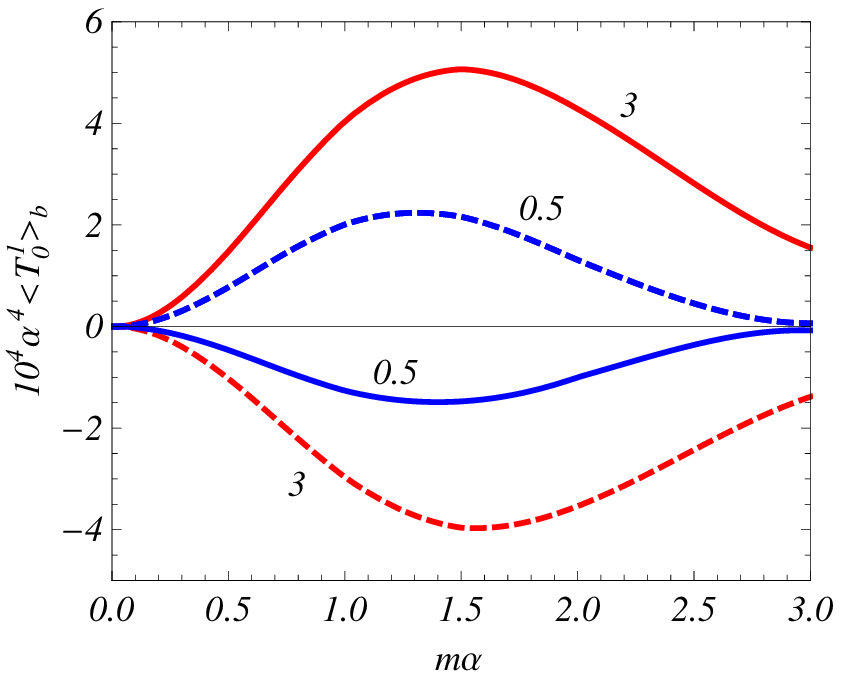,width=7.cm,height=6.cm}%
\end{tabular}%
\end{center}
\caption{Shell-induced contributions in the VEVs of the energy
density (left panel) and the energy flux (right panel) versus the
mass for a $D=3$ conformally coupled field. The graphs are plotted
for $a/\protect\eta =2$ and for fixed values of $r/\protect\eta
=0.5$ and $r/\protect\eta =3$ (numbers near the curves). The
full/dashed curves correspond to Dirichlet/Neumann BCs.}
\label{fig3}
\end{figure}

Figure \ref{fig3} shows the dependence of the boundary-induced parts
in the VEVs of the energy density (left panel) and the energy flux
(right panel) on the mass for a conformally coupled field in $D=3$.
The graphs are plotted for $a/\eta =2$ and for fixed values of
$r/\eta =0.5$ and $r/\eta =3$ (numbers near the curves). The
full/dashed curves correspond to Dirichlet/Neumann boundary
conditions. As it is seen from the presented examples, the VEVs for
massive fields can be essentially larger than those in the massless
case.

\section{Conclusion}

\label{sec:Conc}

In the present paper, for a free scalar field with general curvature
coupling, we have investigated the change in the properties of the
vacuum state induced by a cylindrical shell in background of dS
spacetime. The Robin BC is imposed on the shell, which includes
Dirichlet and Neumann BCs as special cases. We have assumed that the
field is prepared in the Bunch-Davies vacuum state. In the region
inside the shell and for non-Neumann BCs the zero mode is absent and
in this region the Bunch-Davies vacuum is a physically realizable
state for all values of the mass. All the properties of the vacuum
are encoded in two-point functions. As such a function we have taken
the positive-frequency Wightman function. Our method for the
evaluation of this function employs the mode summation and, for the
interior region, is based on a variant of the generalized Abel-Plana
formula. This enabled us to extract explicitly the boundary-free dS
part and to present the contribution induced by the shell in terms
of strongly convergent integrals. The latter
is given by the expression (\ref{WFb}) in the interior region and by (\ref%
{WFbext3}) for the exterior region. In the limit $\alpha \rightarrow \infty $%
, by using the uniform asymptotic expansions for the modified Bessel
functions, the Wightman function is obtained for a shell in Minkowski
spacetime.

Having the Wightman function, we have evaluated the VEVs of the
field squared and of the energy-momentum tensor. These VEVs are
decomposed into boundary-free dS and shell-induced parts. The
boundary-free parts are widely discussed in the literature and we
have concerned with the boundary-induced effects. For points outside
the shell, the shell-induced parts of the VEVs are finite and the
renormalization is reduced to that for the boundary-free geometry.
The shell-induced contributions depend on the variables $\eta $,
$a$, $r$ through the ratios $a/\eta $ and $r/\eta $ which are the
proper radius and the proper distance from the shell axis, measured
in units of the dS curvature scale. This property is a consequence
of the maximal symmetry of dS spacetime and of the Bunch-Davies
vacuum state.

The shell-induced parts in the VEV of the field squared is given by
expressions (\ref{phi2b}) and (\ref{phi2ext}) for the interior and
exterior regions respectively. For $\nu \geqslant 0$ the VEV is
negative for Dirichlet BC and positive for Neumann BC in both
regions. The
boundary-induced part diverges on the shell with the leading term given by (%
\ref{phi2bnear}) for the interior region (in the exterior region
$a-r$ should be replaced by $r-a$). For points near the shell the
effects of the curvature are subdominant and the leading term
coincides with that for the shell in Minkowski bulk. On the axis of
the shell the term $n=0$ contributes only and one has formula
(\ref{phi2Axis}). Simple expressions are obtained for large values
of the shell proper radius compared to the dS curvature scale,
$a/\eta \gg 1$. For positive $\nu $, on the axis, the shell-induced
VEV of the field squared behaves like $(a/\eta )^{2\nu -D}$, whereas
for purely imaginary $\nu $ the corresponding behavior, as a
function of $a/\eta $, is damping oscillatory (see
(\ref{phi2Axis2})). In the exterior region, at proper distances from
the shell larger than the curvature radius of the background
spacetime, the influence of the gravitational field on the
boundary-induced VEVs is crucial. For positive values of $\nu $ and
for non-Neumann BC the shell-induced part in the VEV of the field
squared decays as $(r/\eta )^{2\nu -D}/\ln (r/a)$. For Neumann BC
the decay is stronger, like $(r/\eta )^{2\nu -D-2}$. In the exterior
region and for Robin BC\ with $A,B\neq 0$, the shell contribution in
the mean field squared is positive for points near the shell and
negative at large distances. For purely imaginary values of $\nu $
the behavior of the boundary-induced VEV at large distances is
damping oscillatory and the
leading term is given by (\ref{phi2bFar}) for non-Neumann BC and by (\ref%
{phi2bFarNim}) for Neumann BC. As before, the damping for Neumann BC\ is
faster.

The diagonal components of the shell-induced contribution in the VEV
of the energy-momentum tensor are given by expression (\ref{Tiib})
inside the shell and by (\ref{Tiibext}) outside the shell. In
addition to the diagonal components, the vacuum energy-momentum
tensor has nonzero
off-diagonal component given by the expressions (\ref{T01}) and (\ref%
{T01bext}) for the interior and exterior regions respectively. This
component describes the energy flux along the radial direction and, in
dependence of the parameters, it can by either positive or negative. Note
that unlike to the case of a shell in Minkowski bulk, for dS background the
axial stresses are not equal to the energy density. Near the shell the
leading term in the expansion of the energy density and of the stresses
parallel to the shell is given by (\ref{Tiinear}). For non-conformally
coupled fields these VEVs diverge as the inverse $(D+1)$-th power of the
proper distance from the boundary and near the shell they have the same sign
in the interior and exterior regions. The leading terms for the normal
stress and energy flux are given by (\ref{T11near}) and the corresponding
divergences are weaker. Near the shell, these components have opposite signs
for the interior and exterior regions. For a minimally coupled field the
energy flux near the shell is directed from the shell for Dirichlet BC and
toward the shell for Neumann BC.

On the shell axis and for large values of the shell radius compared to the
dS curvature scale, the diagonal components decay as $(a/\eta )^{2\nu -D}$
for positive values of $\nu $ and exhibit a damping oscillatory behavior,
given by (\ref{Tiiaxis2}), for imaginary $\nu $. The energy flux vanishes on
the axis as $r/\eta $ for $r\rightarrow 0$ and the corresponding asymptotic
expressions are given by (\ref{T01axis}) and (\ref{T10axis1}) for positive
and imaginary $\nu $, respectively. At distances from the shell larger than
the dS curvature scale, to the leading order, the vacuum stresses are
isotropic. For non-Neumann BCs the leading terms in the asymptotic expansion
of the VEV of the energy-momentum tensor are given by (\ref{Tikfar}) and (%
\ref{Tikfarosc}) for positive and imaginary values of $\nu $. In the first
case the diagonal components decay as $(r/\eta )^{2\nu -D}/\ln (r/a)$ and
the asymptotic for the energy flux contains an additional suppression factor
$\eta /r$. For imaginary $\nu $ the decay of the VEVs is oscillatory. This
type of behavior is a gravitationally induced effect and is absent in
Minkowski bulk. For non-Neumann BCs the leading terms in the asymptotic
expansions at large distances do not depend on the specific values of the
coefficients in Robin BC. For minimally and conformally coupled massive
fields and for positive $\nu $, the shell-induced contribution in the energy
density is negative at large distances and the energy flux is directed from
the shell. For Neumann BC, the leading terms in the asymptotic expansion at
large distances are given by expressions (\ref{TikfarN}) and (\ref%
{TikfarNosc}) for positive and imaginary $\nu $ respectively. In this case
the decay of the VEVs at large distances is faster (by an additional factor $%
(\eta /r)^{2}$) than that for non-Neumann BCs. For minimally and conformally
coupled massive fields with Neumann BC and for positive values of $\nu $,
the shell-induced contribution in the energy density is positive and the
energy flux is directed toward the shell. For Robin BC\ with $A,B\neq 0$,
the shell-induced energy density in the exterior region is positive near the
shell and negative at large distances, whereas the energy flux is negative
near the shell and positive at large distances. At some intermediate value
of the radial coordinate these quantities vanish.

\section*{Acknowledgments}

AAS was supported by State Committee Science MES RA, within the frame of the
research project No. SCS 13-1C040.

\end{document}